\documentclass[a4,useAMS,usenatbib,usegraphicx]{mn2e}
\def\apss{\ref@jnl{Ap\&SS}} 
\def\aapr{\ref@jnl{A\&A~Rev.}}
\def\na{\ref@jnl{New~Astronomy}}

\usepackage{amsmath}
\include{epsf}
\usepackage{float}
\usepackage{color}
\usepackage{multirow}
\usepackage{epsf}
\usepackage{url}
\epsfclipon
\title[Exponential disk potential]{Simple and accurate modelling of the gravitational potential produced by thick and thin exponential disks}
\author[R. Smith et~al.]{R. Smith$^{1,2,3}$\thanks{E-mail:rsmith@astro-udec.cl}, C. Flynn$^{4,5}$, G. N. Candlish${^3}$, M. Fellhauer${^3}$, B. K. Gibson${^6}$ \\
$^{1}$Yonsei University, Graduate School of Earth System Sciences-Astronomy-Atmospheric Sciences, Yonsei-ro 50, Seoul 120-749, Republic of Korea \\
$^{2}$CEA-Saclay, DSM, DAPNIA, Service d'Astrophysique, 91191 Gif-sur-Yvette, France\\
$^{3}$Departamento de Astronomia, Universidad de Concepcion, Casilla 160-C, Concepcion, Chile\\
$^{4}$Swinburne University of Technology, PO Box 218, Hawthorn, Victoria 3122, Australia \\
$^{5}$ ARC Centre of Excellence for All-sky Astrophysics, CAASTRO, GPO Box 2702, Canberra, ACT 2601, Australia \\ 
$^{6}$Jeremiah Horrocks Institute, University of Central Lancashire, Preston, PR1~2HE, United Kingdom}
\begin{document}

%\date{Accepted...}
\date{Accepted to MNRAS \today, 7 pages, 7 figures, 2 tables}

\pagerange{\pageref{firstpage}--\pageref{lastpage}} \pubyear{2014}

\maketitle

\label{firstpage}

\begin{abstract}
We present accurate models of the gravitational potential produced by a radially exponential disk mass distribution. The models are produced by combining three separate Miyamoto-Nagai disks. Such models have been used previously to model the disk of the Milky Way, but here we extend this framework to allow its application to disks of any mass, scalelength, and a wide range of thickness from infinitely thin to near spherical (ellipticities from 0 to 0.9). The models have the advantage of simplicity of implementation, and we expect faster run speeds over a double exponential disk treatment. The potentials are fully analytical, and differentiable at all points. The mass distribution of our models deviates from the radial mass distribution of a pure exponential disk by $<$0.4~$\%$ out to 4 disk scalelengths, and $<$1.9~$\%$ out to 10 disk scalelengths. We tabulate fitting parameters which facilitate construction of exponential disks for any scalelength, and a wide range of disk thickness (a user-friendly, web-based interface is also available). Our recipe is well suited for numerical modelling of the tidal effects of a giant disk galaxy on star clusters or dwarf galaxies. We consider three worked examples; the Milky Way thin and thick disk, and a disky dwarf galaxy.
\end{abstract}
\begin{keywords}
methods: numerical --- galaxies: kinematics and dynamics
\end{keywords}

\section{Introduction}

The mass distribution of the stellar disk of most galaxies is well represented by a radially exponential profile (\citealp{Freeman1970}). It is advantageous to be able to accurately model the potential field, accelerations, or tides that arise from a mass distribution with an exponential profile. A radially exponential disk profile has the following form:

\begin{equation}
\label{expdisk}
\Sigma(R) = \Sigma_0 {\rm{exp}} (-R/R_{\mathrm{d}})
\end{equation}

\noindent
where $\Sigma$ is the surface density, $\Sigma_{\rm{0}}$ is central 
surface density, $R$ is radius within the disk, and $R_{\rm{d}}$ is the 
disk scalelength. 

Radially exponential disks may have different vertical density distributions. For galaxy disks, a commonly used form for the vertical density distribution is a ${\rm{sech}}^n$ form:
\begin{equation}
\rho(R,z)=\rho_0 \exp(-R/R_{\mathrm{d}}) {\rm{sech}}^n(-|z|/z_{\mathrm{0}}){\rm{.}}
\label{sechneqn}
\end{equation}

where $z_0$ is the scaleheight, and $n$ is typically $\sim$1 to 3.

Another form of radially exponential disk, the `double exponential', has an exponentially decaying vertical distribution:
\begin{equation}
\rho(R,z)=\rho_0 \exp(-R/R_{\mathrm{d}}) \exp(-|z|/h_{\mathrm{z}}){\rm{.}}
\end{equation}
where $h_{\rm{z}}$ is the exponential disk scaleheight. In fact the double exponential is a special case of Eqn. \ref{sechneqn} when $n$$\rightarrow$$\infty$. To calculate the potential from a double exponential disk, it is necessary to perform the following integral (\citealp{BT1987}):

\begin{equation}
\begin{split}
\Phi(R,z) = - \frac{4 G \Sigma_0}{R_{\rm{d}}} \int^\infty_{- \infty} dz' \exp(-z'/h_z) \times \\ \int^\infty_0 dR' \sin^{-1} \left(\frac{2R'}{A_{\rm{+}}+A_{\rm{-}}}\right) R' K_0(R'/R_d)
\end{split}
\end{equation}
\noindent
where $A_{\rm{+}}=\sqrt{z^2+(R'+R)^2}$, $A_{\rm{-}}=\sqrt{z^2+(R'-R)^2}$ and $K_0$ is the modified Bessel function of the second kind. This integral cannot be performed analytically, and therefore is calculated numerically (e.g \citealp{Dehnen1998}; the {\sc{galpy}} package: \citealp{Bovy2010}), except for in the special case where an infinitely thin exponential disk is assumed. 

Due to these limitations, the potential of disk galaxies has often been modelled using a single Miyamoto-Nagai (MN) disk (e.g \citealp{Allen1991}; \citealp{Fellhauer2006}; \citealp{Fellhauer2007}; \citealp{Kupper2010}; \citealp{Smith2013II}), as this is analytical, and fully defined and provides continuous derivatives at all points. The potential of a single Miyamoto-Nagai (MN) disk is described by the following expression (\citealp{Miyamoto1975}):

\begin{equation}
\Phi(R,z) = \frac{-G M_{\rm{MN}}}{\sqrt{R^2+(a+\sqrt{z^2+b^2})^2}}
\label{MNphi}
\end{equation}
\noindent
where $M_{\mathrm{MN}}$ is the total disk mass, $a$ is the radial scalelength, and $b$ is the vertical scaleheight. This expression can be trivially differentiated in the $R$ and $z$ direction to produce expressions for the acceleration at any location. If $a=0$, the potential of a single MN disk reduces to that of a Plummer distribution (i.e spherical). For $b=0$, the potential reduces to that of an infinitely thin Kuzmin disk (\citealp{kuzmin1956}). Hence by varying the parameters $a$ and $b$, mass distributions can be modelled with a range of thicknesses. Additionally, the density of a single MN disk is given by:
\begin{equation}
\rho(R,z) = \frac{M_{\rm{MN}} b^2 \left[a R^2+(a+3\sqrt{z^2+b^2})(a+\sqrt{z^2+b^2})^2\right]} {4 \pi \left[R^2+(a+\sqrt{z^2+b^2})^2 \right]^{5/2} (z^2+b^2)^{3/2}} 
\label{MNrho}
\end{equation}

While most galaxy disks are well represented by a radially exponential profile (see Eqn. \ref{expdisk}), a single MN disk is a poor match to a radially exponential disk. Its surface density near the centre is too low, and it attains too high surface density at large radius. We quantify the deviation from a pure exponential disk for the models considered in this paper in the following manner. We calculate the difference in mass found within a radial annuli of a single MN disk and a pure exponential disk, and sum up the absolute of these differences in all annuli out to a chosen radius. In this way, we calculate that the total mass deviation of a single, thin, MN disk from a pure exponential disk is 5.0$\%$ at 4 $R_{\mathrm d}$, and 19.9$\%$ at 10 $R_{\mathrm d}$. 

To improve the match to a radially exponential disk, \cite{Flynn1996} combined three MN (3MN) disk profiles, each with different radial scalelength $a$, one with negative mass, and all with a single vertical scaleheight $b$. This model is better at matching a radially exponential disk at large radii, and we calculate the total mass deviation from a thin, radially exponential disk is 9.4$\%$ at 4 $R_{\mathrm d}$, and 10.0$\%$ at 10 $R_{\mathrm d}$. This 3MN model has since been used extensively for modelling the potential from the disk of the Milky Way (e.g \citealp{Hanninen2004}; \citealp{Rodionov2008}; \citealp{Moni2014}). However minor alterations to the original parameter values have been used to better match more recent measurements of the Milky Way's disk scalelength and circular velocity (\citealp{Gardner2010}; \citealp{Jilkova2012}; \citealp{Loyola2014}).

The aim of this study is to extend the utility of the framework introduced by \cite{Flynn1996}, a framework which was developed specifically for the recovery of an Milky Way-like radially exponential disk using 3MN potentials. Our extension here allows its general application to disks of any mass, scalelength, and thickness, rather than simply for the singular purpose for which it was initially designed. Our new models also better match the distribution of a radially exponential disk. In $\oint$\ref{New3MNmodels} we derive the new 3MN models, in $\oint$\ref{Workedexamples} we consider three worked examples, in $\oint$\ref{thickdiskslices} we compare a 3MN model to other well known disk models, and finally we summarise and conclude in $\oint$\ref{Conclusions}.

\section{New triple MN (3MN) potentials}
\label{New3MNmodels}
From Eqn. \ref{MNphi}, a single MN potential has 3 free parameters; disk mass ($M_{\mathrm d}$), radial scalelength ($a$), and vertical scaleheight ($b$). Following \cite{Flynn1996}, we choose a single fixed value of $b$ for all three MN potentials. This substantially reduces the total parameter space we must consider, and ensures our models have a uniform scaleheight. It also enables us to control disk thickness through a single parameter. Thus our 3MN models have a total of 7 free parameters; $M_{\mathrm{MN,1}}$, $M_{\mathrm{MN,2}}$, $M_{\mathrm{MN,3}}$, $a_1$, $a_2$, $a_3$, and $b$. Our aim is to find a combination of these 7 parameters that minimises the mass deviation from a radially exponential disk, out to 4 $R_{\mathrm d}$\footnote{The choice of 4 $R_{\mathrm d}$ is rather arbitrary, however $>$90$\%$ of an exponential disk mass is enclosed within 4 $R_{\mathrm d}$}.

In order to find a good combination we used a brute force approach. We fix the value of $b$, then numerically ran through a grid of values for the other 6 parameters $M_{\mathrm{MN,1}}$, $M_{\mathrm{MN,2}}$, $M_{\mathrm{MN,3}}$, $a_1$, $a_2$, $a_3$. For each parameter set, the mass deviation from a pure exponential disk out to 4 and 10 $R_{\mathrm d}$ was quantified numerically. Due to the significant numbers of combinations of parameter values possible, we initially ran with a coarse grid, and then later ran with a finer grid focussed on the best matches.

Using this brute-force approach, we find that for an infinitely thin disk ($b$/$R_{\mathrm d}$=0.0), {\it{a 3MN model can be found that radially deviates from a radially exponential disk by only 0.4$\%$ within 4 $R_{\mathrm d}$, and only 1.9$\%$ out to 10 $R_{\mathrm d}$.}} 

\subsection{A recipe for different disks: varying disk mass, size and thickness}
\label{diskthickness}
Our choice of 3MN model has a highly useful property -- once a good match to a radially exponential disk of mass $M_{\mathrm d}$ and scalelength $R_{\mathrm d}$ has been found, it can be easily scaled to different masses and disk sizes. For example, an equally good match can be found for a disk twice as massive by simply doubling all the MN disk masses ($M_{\mathrm{MN,1}}$, $M_{\mathrm{MN,2}}$, $M_{\mathrm{MN,3}}$).

\begin{figure} \centering \epsfysize=8.0cm \epsffile{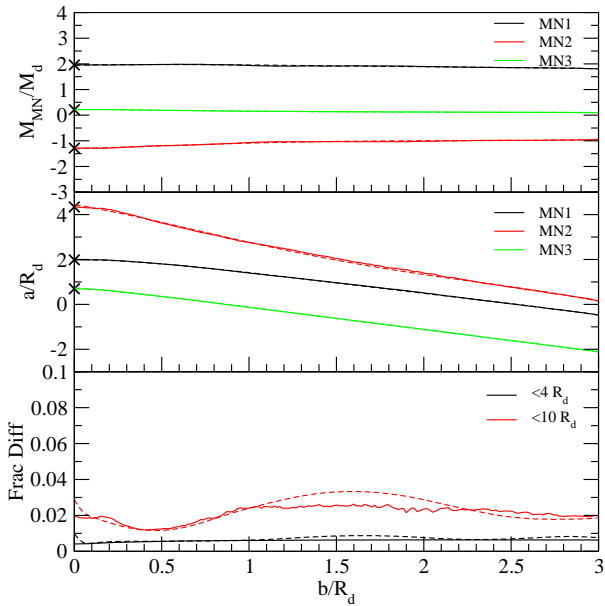} 
\caption{(Top panel) Evolution of the three mass parameters, and (middle panel) three scalelength parameters of the 3MN model, as a function of disk thickness $b$/$R_{\mathrm d}$. Cross symbols are the solution from the brute-force approach for an infinitely thin disk. Solid lines are the continous solutions for a range of disk thicknesses. Dashed lines are 4th-order fits to the solid lines. (Lower panel) The fractional difference in mass between the model and a pure exponential disk measured radially outwards to 4 scalelengths (black), and 10 scalelengths (red).}
\label{6paramset}
\end{figure}

Therefore to calculate the best 3MN model with a different thickness, a new 6-parameter set must be found. However, if the change in $b$/$R_{\mathrm d}$ is very small, and smooth, we can expect that each parameter changes in a smooth, and continuous manner. We initially searched for best matches for different disk thicknesses using the brute force approach. However in practice it was difficult to find a smooth fit-line through these points for which every point on the fit-line provided a good radial match to an exponential. This was also highly time consuming as the brute force approach involves searching through all possible combinations of a broad range of values for each of the 6 parameters. Thus we changed our approach -- instead we use our best solution for an infinitely thin disk, found by the brute-force approach, as a prior, and then use an alternative approach to find the variation of each parameter as a function of $b$/$R_{\mathrm d}$. 

\begin{table}
\centering
\begin{tabular}{|c|c|c|c|c|c|}
Parameter & $k_1$ & $k_2$ & $k_3$ & $k_4$ & $k_5$ \\
\hline \hline
$M_{\mathrm{MN,1}}/M_{\mathrm{d}}$ & $-$0.0090 &  0.0640 & $-$0.1653 &  0.1164 &  1.9487 \\
$M_{\mathrm{MN,2}}/M_{\mathrm{d}}$ &  0.0173 & $-$0.0903 &  0.0877 &  0.2029 & $-$1.3077 \\
$M_{\mathrm{MN,3}}/M_{\mathrm{d}}$ & $-$0.0051 &  0.0287 & $-$0.0361 & $-$0.0544 &  0.2242 \\
$a_{\mathrm{1}}/R_{\mathrm{d}}$    & $-$0.0358 &  0.2610 & $-$0.6987 & $-$0.1193 &  2.0074 \\
$a_{\mathrm{2}}/R_{\mathrm{d}}$    & $-$0.0830 &  0.4992 & $-$0.7967 & $-$1.2966 &  4.4441 \\
$a_{\mathrm{3}}/R_{\mathrm{d}}$    & $-$0.0247 &  0.1718 & $-$0.4124 & $-$0.5944 &  0.7333 \\
\end{tabular}
\caption{Table of constants in Eqn. \ref{4thorder}, providing an accurate fit to the variation of each of the 6 parameters (see column 1) of the 3MN model shown in Fig. \ref{6paramset}.}
\label{4thorderterms}
\end{table}

See upper and central panel of Fig. \ref{6paramset}. The best solution from the brute force approach is shown by cross symbols. We then shift along the $b$/$R_{\mathrm d}$ axis in small steps, searching for new solutions that allow a continuous path across the figure, while simultaneously attempting to minimise the differences between the 3MN model and that of an exponential. In practice this approach is significantly less computationally challenging in comparison to the brute-force approach, as only a small range of possible values for each parameter are permitted in order to form a continuous path. As a result, in a fraction of the computational time required for the brute-force approach to find a single parameter set, hundreds of parameter set solutions are found for a wide range of $b$/$R_{\mathrm d}$ value. These solutions naturally form a continuous path, allowing a user to choose an arbitrary value of $b$/$R_{\mathrm d}$ within the permitted range. The solutions are shown by the solid lines in Fig. \ref{6paramset}, and cover a range of disk thickness from $b$/$R_{\mathrm d}$=0.0 to 3.0.

The dashed lines in the upper and central panel show a 4th-order fit to the solid lines. {\it{This fit produces 3MN models that match a radially exponential disk to $<$$1.0\%$ out to 4 $R_{\mathrm d}$, and $<$$3.3\%$ out to 10 $R_{\mathrm d}$ (dashed black and red line in lower panel respectively), for the range of disk thickness considered.}} The 4th-order fit to each parameter has the form:

\begin{figure} \centering \epsfysize=8.0cm \epsffile{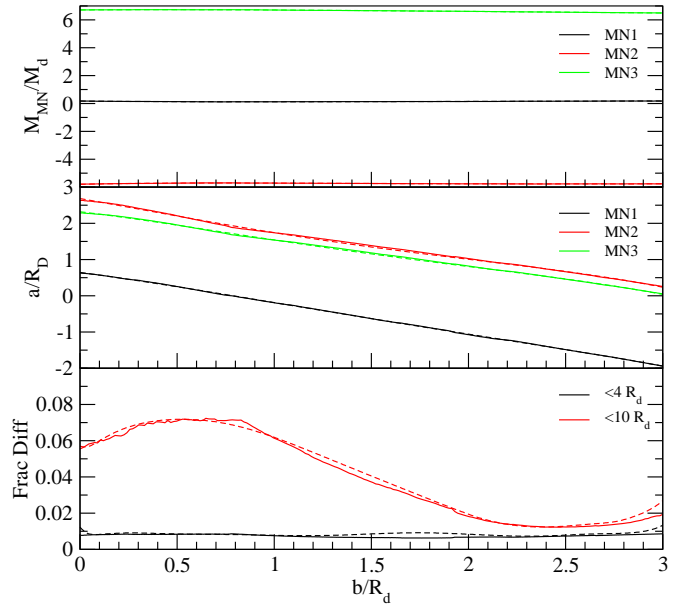} 
\caption{Caption as in Fig \ref{6paramset}, except now for models with positive densities at all positions for the thickness range $b$/$R_{\mathrm d}$=0 to 1.35.}
\label{6paramsetII}
\end{figure}

\begin{table}
\centering
\begin{tabular}{|c|c|c|c|c|c|}
Parameter & $k_1$ & $k_2$ & $k_3$ & $k_4$ & $k_5$ \\
\hline \hline
$M_{\mathrm{MN,1}}/M_{\mathrm{d}}$ &  0.0036 & $-$0.0330 &  0.1117 & $-$0.1335 &  0.1749 \\
$M_{\mathrm{MN,2}}/M_{\mathrm{d}}$ & $-$0.0131 &  0.1090 & $-$0.3035 &  0.2921 & $-$5.7976 \\
$M_{\mathrm{MN,3}}/M_{\mathrm{d}}$ & $-$0.0048 &  0.0454 & $-$0.1425 &  0.1012 &  6.7120 \\
$a_{\mathrm{1}}/R_{\mathrm{d}}$    & $-$0.0158 &  0.0993 & $-$0.2070 & $-$0.7089 &  0.6445 \\
$a_{\mathrm{2}}/R_{\mathrm{d}}$    & $-$0.0319 &  0.1514 & $-$0.1279 & $-$0.9325 &  2.6836 \\
$a_{\mathrm{3}}/R_{\mathrm{d}}$    & $-$0.0326 &  0.1816 & $-$0.2943 & $-$0.6329 &  2.3193 \\
\end{tabular}
\caption{Table of constants in Eqn. \ref{4thorder}, providing an accurate fit to the variation of each of the 6 parameters of the 3MN model shown in Fig. \ref{6paramsetII}. These models have positive densities at all positions for the disk thickness range $b$/$R_{\mathrm d}$ from 0 to 1.35.}
\label{newtab}
\end{table}

\begin{equation}
{\rm{Parameter}} = k_1 x^4 + k_2 x^3 + k_3 x^2  + k_4 x + k_5
\label{4thorder}
\end{equation}
\noindent
where $x=b$/$R_{\mathrm d}$. The values of the constants ($k_1$-$k_5$) are given in Tab. \ref{4thorderterms}.  

\begin{figure} \centering \epsfysize=6.0cm \epsffile{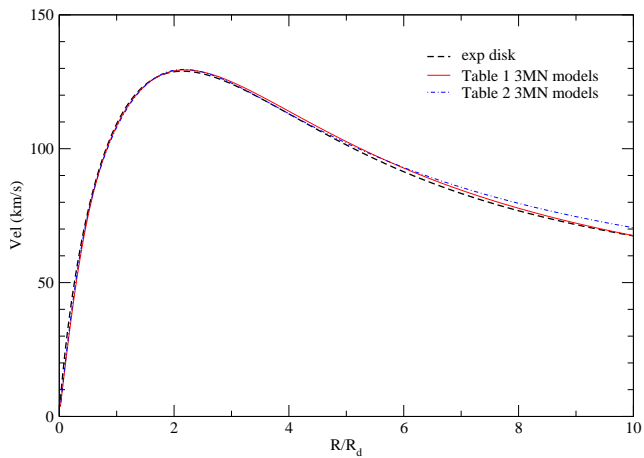} 
\caption{Rotation curve of an infinitely thin exponential disk with $M_{\mathrm d}$=1.0$\times$10$^{10}$~M$_\odot$, and $R_{\mathrm d}$=1.0~kpc (black dashed line). For comparison, the rotation curve of the 3MN models from Tab. \ref{4thorderterms} is shown (red solid line), and the rotation curve of the 3MN models from Tab \ref{newtab} is shown (blue dash-dot line). Within 4 scalelengths, all models are barely distinguishable.}
\label{vcirccurves}
\end{figure}

\subsection{Issues with negative densities}
A shortcoming with the models described in Fig. \ref{6paramset} and Tab. \ref{4thorderterms} are that the MN model with negative mass, also has the largest scalelength. As a result it is inevitable that at sufficiently large radius, negative densities will be found. Negative densities at large radii also ocurred in the original \cite{Flynn1996} model. In practice, the negative densities occur near the plane of the disk, and only in the very outer disk (at $R=$5.6 to 11.2 scalelengths for $b$/$R_{\mathrm d}$ in the range 0.0 to 3.0). As a result the negative densities are very small, and if the disk is placed within a dark matter halo, then the negative densities will be more than offset by the halo leaving only positive densities everywhere. If this is of concern (e.g if disks without halos are studied), we also present alternative models. These models are advantageous as they have positive densities at all positions, for disk thickness $b$/$R_{\mathrm d}$ from 0 to 1.35. For $b$/$R_{\mathrm d}>$$1.35$, negative densities begin to appear in the outer disk of these models as well. The lower panel of Fig. \ref{6paramsetII} shows that these models are still very accurate out to 4 scalelengths (solid black line): $<$1$\%$ difference from a pure exponential). However they are often less accurate out to 10 scalelengths (solid red line): 1.5$\%$ to 7$\%$ difference from a pure exponential, with the value being quite sensitive to disk thickness. Once again the dashed lines show a 4th order fit of the form shown in Eqn. \ref{4thorder}. We provide the values of the constants $k_1$-$k_5$ for these additional models in Tab. \ref{newtab}. Although we also found other models with positive densities at all points for disks as thick as $b$/R$_{\rm{d}}\sim2.0$, these models were significantly less accurate matches to radially exponential disks, and we will not consider them further.

\subsection{Rotation curves with the 3MN models}
In Fig. \ref{vcirccurves}, we compare the rotation curve for an infinitely thin exponential disk (black dashed line) to those of the 3MN models from Tab. \ref{4thorderterms} (red solid line) and from Tab. \ref{newtab} (blue dot-dash line). Within 4 scalelengths, the rotation curves are all virtually indistinguishable. At large radius the greater accuracy of the models in Tab. \ref{4thorderterms} is visible in comparison to the Tab. \ref{newtab} models. We note that the negative densities arising in the Tab. \ref{4thorderterms} models, appear in the plane of the disk no closer than 8.0 scalelengths for an infinitely thin model. They have negligible effect on the rotation curve.

\subsection{Useful conversion formulae}
\subsubsection{Disk thickness ($b$/$R_{\mathrm d}$) and ellipticity ($e$)}
\begin{figure} \centering \epsfysize=6.0cm \epsffile{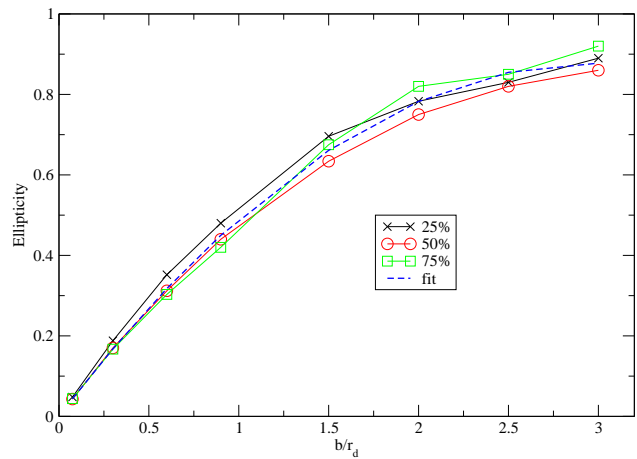} 
\caption{Conversion from disk thickness ($b$/$R_{\mathrm d}$) to ellipticity for 3MN models viewed edge-on, and with ellipticity measured for contours containing 25$\%$ (black), 50$\%$ (red), and 75$\%$ (green) of the total mass. The disk ellipticity is found to be roughly equal for all three contours. A quadratic fit is made to all the data points (blue dashed line) and the fit is given in Eqn. \ref{converte}.}
\label{thicknessmatch}
\end{figure}

To aid comparison with observed disk ellipticities, we measure how the ellipticity $e$ of our disk models, seen edge-on, varies with disk thickness $b$/$R_{\mathrm d}$. We measure the ellipticity of contours containing 25$\%$, 50$\%$, and 75$\%$ of the total mass, for a range of disk thickness from $b$/$R_{\mathrm d}=0$ to 3. The results are shown in Fig. \ref{thicknessmatch}. In practice, the ellipticity changes very little whether we use the contour containing 25$\%$, 50$\%$ or 75$\%$ of the total mass. This indicates the ellipticity of the 3MN model is only a weak function of radius. We fit the data points with a quadratic formula, shown by the blue dashed line in Fig. \ref{thicknessmatch}, giving ellipticity $e$ as a function of disk thickness $b$/$R_{\mathrm d}$:

\begin{equation}
\label{converte}
e = -0.099 (b/R_{\rm{d}})^2 + 0.599 (b/R_{\rm{d}}){\rm{.}}
\end{equation}

Solving Eqn. \ref{converte}, we obtain an expression for converting from disk thickness $b$/$R_{\mathrm d}$ as a function of ellipticity $e$:

\begin{equation}
\label{convertthick}
b/R_{\rm{d}} = 3.025 - 3.178 \sqrt{0.906-e}{\rm{.}}
\end{equation}

\begin{figure} \centering \epsfysize=4.2cm \epsffile{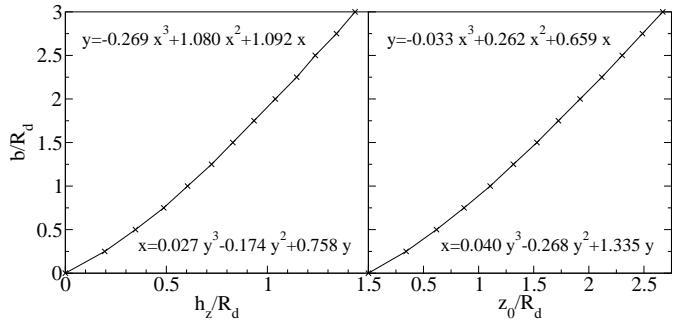} 
\caption{Conversions between 3MN disk thickness ($b$/$R_{\mathrm d}$) and (left panel) exponential scaleheight, or (right panel) sech$^2$ scaleheight. Data points are found using the technique described in Sect. \ref{hzconvert}. The equations provide smooth fits through the data points. In each panel, the upper-left equation gives conversion to $b$/$R_{\mathrm d}$, and the lower-right equation gives conversion from $b$/$R_{\mathrm d}$. Conversions are approximate, but are best matches.}
\label{bconvert}
\end{figure}

\subsubsection{Disk thickness to exponential and sech$^2$ scaleheight}
\label{hzconvert}

Our 3MN models have a single parameter controlling their thickness, $b$/$R_{\mathrm d}$. However users are likely to be unfamiliar with $b$/$R_{\mathrm d}$ as a measure of thickness. For this reason, we characterise this to scalelengths for more familiar vertical density distributions; an exponential, or a sech$^2$ decay with $z$.

We first sum up the fractional difference between the density of the 3MN model and the exponential, measured vertically out of the plane up to 5 scaleheights ($z$=0 to 5$b$), measured at $R$=0. We vary the disk thickness $b$/$R_{\mathrm d}$ in order to minimise the sum, and best match the exponential distribution. We then tabulate the best matches we find between $b$/$R_{\mathrm d}$ and $h_{\mathrm z}$/$R_{\mathrm d}$ over a range of $b$/$R_{\mathrm d}$ from 0 to 3. We repeat this procedure to find the best match between 3MN disks and a sech$^2$ vertical distribution. The best match is the same if we instead choose to compare the vertical density profiles at  $R=2$ or 4~$R_{\mathrm d}$. 

The tabulated best matches are plotted in Fig. \ref{bconvert}. The given formulae provide smooth fits through the tabulated data points. In each panel, the upper-left equation provides conversion to $b$/$R_{\mathrm d}$, while the lower-right equation provides conversion from $b$/$R_{\mathrm d}$. We emphasise that our 3MN models are mathematically ill-equipped to tightly match an exponential or sech$^2$ vertical density distribution at all $z$ (an example will be shown in Sect. \ref{thickdiskslices}). Thus the purpose of these equations is only to provide users with the approximate (but best available) match. This in turn provides a sense of the physical meaning of a particular choice of $b$/$R_{\mathrm d}$ value.

\section{Worked examples}
\label{Workedexamples}
\begin{figure} \centering \epsfysize=12.5cm \epsffile{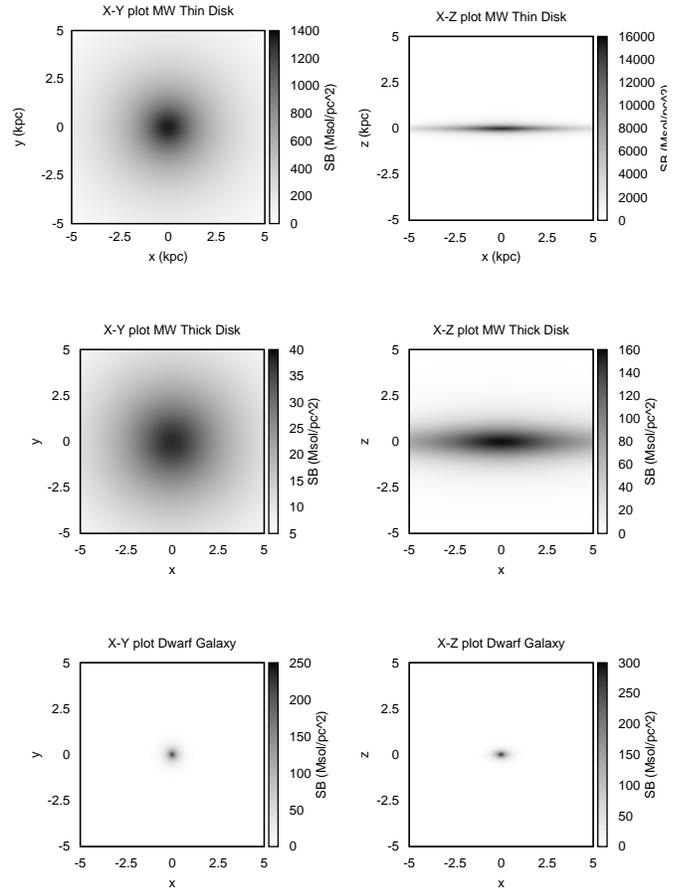} 
\caption{Projected surface density plots for (upper) the thin disk model, centre the thick disk model, (lower) the dwarf galaxy model. The left column is face-on projection, and the right column is edge-on projection. The grey scale bar gives surface densities in units of M$_\odot$~pc$^{-2}$. All panels are 10 kpc on a side, to give a sense of relative size.}
\label{projplots}
\end{figure}

Finally we consider three worked examples; the thick disk of a disky dwarf galaxy, the thin disk of the Milky Way (MW), and the thick disk of the MW. Projected density plots for all three models are shown in Fig. \ref{projplots}. For these examples we use the more accurate Tab. \ref{4thorderterms} models.

\subsection{A dwarf disk galaxy with a thick disk}
We consider a dwarf disk with a total disk mass $M_{\mathrm d}=1\times10^8$~M$_\odot$, and $R_{\mathrm d}=0.25$~kpc (\citealp{Fathi2010}). We assume an ellipticity of 0.6 (\citealp{Ruben2010}). From Eqn. \ref{convertthick}, $b/R_{\mathrm d}=1.27$. Substituting $b/R_{\mathrm d}$ into Eqn. \ref{4thorder} with constants from Tab. \ref{4thorderterms} we get; $M_{\mathrm{MN,1}}=1.94 \times 10^{8}$~M$_\odot$, $M_{\mathrm{MN,2}}=-1.05 \times 10^{8}$~M$_\odot$, $M_{\mathrm{MN,3}}=0.142 \times 10^{8}$~M$_\odot$, $a_{\mathrm{1}}=0.29$~kpc, $a_{\mathrm{2}}=0.58$~kpc, $a_{\mathrm{3}}=-0.10$~kpc. This model matches a radially exponential disk to better than 0.8$\%$ at 4 $R_{\mathrm d}$, and 3.0$\%$ at 10 $R_{\mathrm d}$. 

\subsection{The thin disk of the MW}
We consider a radially exponential disk for the MW's thin disk with a total mass of $M_{\mathrm d}=4.6\times$10$^{10}$~M$_\odot$, and radial exponential scalelength $R_{\mathrm d}=2.2$~kpc (\citealp{Bovy2013}).  We choose an exponential scaleheight of $h_z=0.2$~kpc (\citealp{Larsen2003}), and Fig. \ref{bconvert} gives $b$/$R_{\mathrm d}=0.11$. Substituting $b$/$R_{\mathrm d}$ into Eqn. \ref{4thorder} with constants from Tab. \ref{4thorderterms}, we get $M_{\mathrm{MN,1}}=9.01 \times 10^{10}$~M$_\odot$, $M_{\mathrm{MN,2}}=-5.91 \times$ 10$^{10}$~M$_\odot$, $M_{\mathrm{MN,3}}=1.00 \times 10^{10}$~M$_\odot$, $a_{\mathrm{1}}=4.27$~kpc, $a_{\mathrm{2}}=9.23$~kpc, $a_{\mathrm{3}}=1.43$~kpc. This model matches a radially exponential disk to better than 0.5$\%$ at 4 $R_{\mathrm d}$, and to better than 1.8$\%$ at 10 $R_{\mathrm d}$.

\begin{figure} \centering \epsfxsize=9.1cm \epsffile{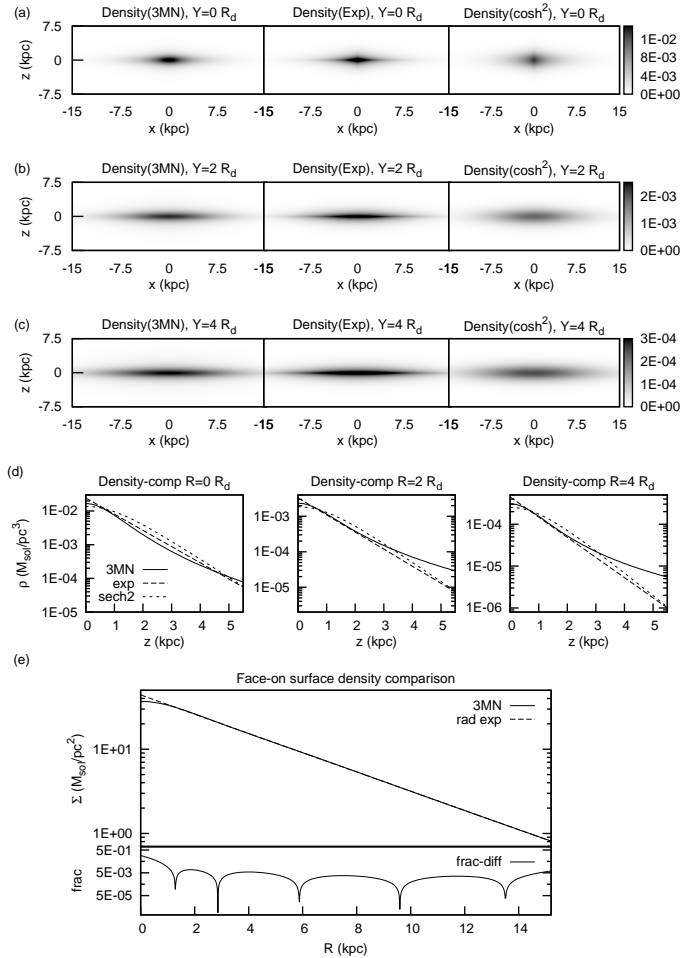} 
\caption{Analysis of the MW thick disk model of Sect. \ref{thickdiskMW}. Rows (a), (b), and (c) are cross-sections through the volume density distribution, and each row allows comparison between the 3MN model (left), a double exponential (middle), and a radially exponential model with a sech$^2$ density drop off out of the plane (right). Grey scale color bar units are $M_\odot~pc^{-3}$. The plane of the disk lies on the $x$-$y$ plane, and slices are perpendicular to the plane at (a) $y$=0$R_\mathrm{d}$, (b) $y$=2$R_\mathrm{d}$, and (c) $y$=4$R_\mathrm{d}$. Row (d) is volume density with distance from the disk plane, measured at (left) 0$R_\mathrm{d}$, (centre) 2$R_\mathrm{d}$, and (right) 4$R_\mathrm{d}$. Different curves (indicated in key) compare the 3MN model, double exponential, and sech$^2$ model. The upper panel of row (e) compares the surface density profile of the 3MN model (solid curve) with a radially exponential disk (dashed curve) out to 4$R_\mathrm{d}$. The lower panel of row (e) shows the absolute fractional difference in the surface density of the two profiles at each radius.}
\label{sliceplots}
\end{figure}

\subsection{The thick disk of the MW}
\label{thickdiskMW}
For the thick disk, we consider a radially exponential disk with scalelength $R_{\mathrm d}=3.8$~kpc and scaleheight $h_z=0.9$~kpc (\citealp{Moni2012}). Using Fig. \ref{bconvert} gives $b/R_{\mathrm d}=0.30$. We assume that the thick disk mass is 8.6$\%$ of the thin disk mass (\citealp{Yoachim2006}) so the total disk mass $M_{\mathrm d}=4.0\times10^{9}$~M$_\odot$. Substituting $b/R_{\mathrm d}$ into Eqn. \ref{4thorder} with constants from Tab. \ref{4thorderterms} gives; $M_{\mathrm{MN,1}}=7.88\times$ 10$^{9}$~M$_\odot$, $M_{\mathrm{MN,2}}=-4.97 \times$ 10$^{9}$~M$_\odot$, $M_{\mathrm{MN,3}}=0.82 \times$ 10$^{9}$~M$_\odot$, $a_{\mathrm{1}}=7.30$~kpc, $a_{\mathrm{2}}=15.25$~kpc, $a_{\mathrm{3}}=2.02$~kpc. This model matches a radially exponential disk to better than 0.6$\%$ at 4 $R_{\mathrm d}$, and to better than 1.4$\%$ at 10 $R_{\mathrm d}$.

\section{Comparison of 3MN model with other density profiles}
\label{thickdiskslices}
In Fig. \ref{sliceplots}, we analyse the 3MN thick disk model from Sect. \ref{thickdiskMW} in more detail. In row (a)-(c) we compare cross-sections through the volume density distribution of the 3MN model (left), a double exponential (middle), and a radially exponential model whose density profile decays as sech$^2$ out of the plane (right). The scaleheight of the exponential and sech$^2$ distribution is the best match to the 3MN model, calculated using the equations in Fig. \ref{bconvert}. Clearly it is impossible for the 3MN model to exactly match the other profiles, as they are mathematically distinct. However, by comparing them we can judge how they differ, and the quality of the best matches provided in Fig. \ref{bconvert}. Comparing along each row, the exponential disk is most `cuspy' in the vertical direction, where as the sech$^2$ is the most `cored', and the 3MN model is found somewhere in between the other two. This is true whether the cross-section is made at (a) $y$=0$R_\mathrm{d}$, (b) $y$=2$R_\mathrm{d}$, or (c) $y$=4$R_\mathrm{d}$.

Row (d) provides a more quantitative analysis of the vertical density distribution up to $\sim$5$b$, measured at (left) $R$=0$R_\mathrm{d}$, (centre) $R$=2$R_\mathrm{d}$, and (right) $R$=4$R_\mathrm{d}$. At small $z$ (less than roughly 2 or 3$b$) the 3MN density distribution roughly matches the other two profiles. However at greater distances from the plane, the 3MN model returns higher densities, and this becomes stronger if the vertical density profile is measured at larger R.

The upper panel of row (e) compares the surface density of the 3MN model (solid curve) to a radially exponential disk (dashed curve). Curves are plotted out to $R$=4$R_\mathrm{d}$. There is clearly an excellent agreement over this radius range, although it can be seen that the 3MN model returns slightly lower densities at very small radii ($R$$<$0.2$R_\mathrm{d}$). The lower panel indicates the absolute fractional difference in surface density between the two profiles. The maximum density difference is $\sim$15$\%$ at $R$=0, but for $R$$>$0.2$R_\mathrm{d}$ the density difference is very low (typically $\sim$$0.5\%$).

\section{Summary and Conclusions}
\label{Conclusions}
We present a recipe for using a triple Miyamoto-Nagai (3MN) disk distribution to model the potential of a radially exponential disk. 3MN models have previously been used to model the disk component of the Milky Way. Here we extend on this framework to allow its general application to disks of any mass, scalelength, and a wide range of thickness. We find parameters of the 3MN model that best match the mass distribution of a radially exponential disk. We consider a broad range of disk thicknesses from infinitely thin (ellipticity $=0.0$), to near spherical (ellipticity $=0.9$). The 3MN models have many benefits as they are entirely analytical, easy to implement, and provide continuous derivatives (enabling a calculation of accelerations) at all points.

\begin{enumerate}
\item We provide accurate fitting formulae to our new 3MN models, that reproduce the mass distribution of a radially exponential disk to $<$$1.0\%$ out to $4 R_{\mathrm d}$, and $<$$3.3\%$ out to $10 R_{\mathrm d}$ for disks with a range of ellipticities from flat to near spherical (see Eqn. \ref{4thorder} and Tab. \ref{4thorderterms}).
\item We provide a second set of models in Tab. \ref{newtab} that ensures positive densities at all positions for the disk thickness range $b$/$R_{\mathrm d}=0.00$ to 1.35. This is equivalent to an ellipticity range from 0.0 to 0.6. 
\item We provide a fitting formula to allow for easy conversion between the disk thickness $b$/$R_{\mathrm d}$ and disk ellipticity $e$ (see Eqn. \ref{converte}), and in reverse (see Eqn. \ref{convertthick}). 
\item The vertical distribution of our 3MN models is similar to an exponential or sech$^2$ distribution at small $z$. We provide a rough approximation for converting between disk thickness, exponential scaleheight, and sech$^2$ scaleheight for $z$ up to 5 scaleheights (see equations given in Fig. \ref{bconvert}).
\end{enumerate}

A user-friendly, online web-form is available at \url{http://astronomy.swin.edu.au/~cflynn/expmaker.php}. Users can request the disk thickness they require. The page will automatically provide the best matching 3MN parameters, calculated using our scheme. We acknowledge that better 3MN solutions may exist, hidden in the parameter space. In the future we will extend on the techniques developed in this study in order to uncover alternative and potentially more accurate 3MN parameter sets, but also to allow for flared disks, and for alternative vertical density distributions.

\section*{Acknowledgements}
CF acknowledges support by the Beckwith Trust. MF acknowledges support by FONDECYT grant 1130521. RS acknowledges support from Brain Korea 21 Plus Program (21A20131500002) and the Doyak Grant(2014003730). RS also acknowledges support support from the EC through an ERC grant StG-257720, and Fondecyt (project number 3120135). GC acknowledges support by FONDECYT grant 3130480. BKG acknowledges the support of the UK Science \& Technology Facilities Council (ST/J001341/1), and the generous visitor support provided by FONDECYT and the Universidad de Concepcion.

\bibliography{bibfile}

\bsp

\label{lastpage}

\end{document}